\documentclass[aps, prl, superscriptaddress, groupedaddress, twocolumn, floatfix, 10pt]{revtex4} 
\usepackage{dcolumn}   
\usepackage{bm}        
\usepackage{amssymb, amsmath}   

\usepackage{graphicx}
\usepackage{tikz}
\usepackage{tikz-3dplot}

\definecolor{gray1}{gray}{0.6}
\definecolor{gray2}{gray}{0.8}

\begin{document}

\newcommand{\abs}[1]{\lvert #1 \rvert}
\newcommand{\note}[1]{\textcolor{red}{\textit{#1}}}

\title{Comment on ``Metal-Insulator Transition in an Aperiodic Ladder Network: An Exact Result''}

\author{Sergej Flach}
\affiliation{New Zealand Institute for Advanced Study, Centre for Theoretical Chemistry \& Physics, Massey University,  0745 Auckland, New Zealand}
\author{Carlo Danieli}
\affiliation{New Zealand Institute for Advanced Study, Centre for Theoretical Chemistry \& Physics, Massey University,  0745 Auckland, New Zealand}
\maketitle

Sil, Maiti, and Chakrabarti (SMC) \cite{sil2008} introduce an aperiodic two-leg ladder network composed of atomic sites with on-site potentials distributed according to a quasiperiodic Aubry-Andre potential.  SMC claim the existence of multiple mobility edges, i.e. metal-insulator transitions at multiple values of the Fermi energy.
SMC use numerical calculations of the conductance and density of states, and an analytical result in a limiting case. In the following, we restate in the real space the change of basis done by SMC. This change of basis reduces their model to two \emph{decoupled} chains, each with its own Aubry-Andre potential.
Each of the chains has its own critical point, where \emph{all} states change from metallic to insulating, without any notion of a mobility edge. Since the two chains are not interacting, their spectra may overlap such that localized states (from one chain) and delocalized states (from the other chain) coexist at the same energy. Consequently, the analysis of
the density of states and the conductance alone is misleading, and that is due to the choice of the symmetric quasiperiodic potential. We briefly outline the possibility to obtain true mobility edges.

Recalling the variables $f_{n,1},f_{n,2}$ from Eq.(4) in \cite{sil2008},
the real space amplitude equations of SMC are given by
\begin{align}
(E - \epsilon_{n,1}  ) f_{n,1} =  \hspace*{5cm}~&
\nonumber
\\
t_l(f_{n-1,1}+f_{n+1,1}) +t_d(f_{n-1,2}+f_{n+1,2}) +\gamma f_{n,2} , \\
(E -\epsilon_{n,2} ) f_{n,2} =\hspace*{5cm}~&
\nonumber
\\
t_l(f_{n-1,2}+f_{n+1,2}) +t_d(f_{n-1,1}+f_{n+1,1}) +\gamma f_{n,1}
\end{align}
where  $E$ is the eigenenergy and $\epsilon_{n,1},\epsilon_{n,2}$ the aperiodic potential in both sublattices.
SMC chose a symmetric potential $\epsilon_{n,1}=\epsilon_{n,2}\equiv \epsilon_n$.
That choice has consequencies.
We rewrite the real space amplitude equations by transforming
to $f_n^{\pm} = (f_{n,1} \pm f_{n,2} ) /\sqrt{2}$:
\begin{align}
[E + \gamma - \epsilon_n] f_n^- &= (t_l - t_d) (f_{n-1}^- + f_{n+1}^- ), \\
[E - \gamma - \epsilon_n ] f_n^+ &= (t_l + t_d ) (f_{n-1}^+ + f_{n+1}^+ ), 
\end{align}
The problem detangles into two independent one-dimensional tight-binding chains with spectra $E^{\pm}$.
For instance for the particular choice of SMC $\epsilon_{n,1}=\epsilon_{n,2}=\lambda \cos (Q n a)$, $a=1$ and $Q$ an irrational
multiple of $\pi$,
both chains correspond to the Aubry-Andre model \cite{aa} which has a metal-insulator transition \emph{without}
any mobility edge. Indeed, all states of any of the two $f^{\pm}$-chains transit simultaneously at $\lambda^{\pm}_c = 2|(t_l\pm t_d)|$. 

Since the transitions happen in general at different values of $\lambda$ for the two different chains, and the chain spectra $E^{\pm}$ are shifted relative to each other for $2\gamma \neq 0$, one may encounter at the same energy a localized state from one
chain, and a delocalized state from the other chain. This is however \emph{not} due to a mobility edge, but simply due
to the artificial overlap of spectra of two noninteracting subsystems. The combined numerical use of the density of states and the conductance does not easily differentiate between the case of a noninteracting and an interacting chain
case, therefore the numerics in SMC leads to false conclusions. The analytical attempt of SMC (as well as the
numerical data in Fig.2 in \cite{sil2008}) to obtain a mobility edge
for $t_l=t_g$ is obviously failing, since in that case the $f_n^-$-chain becomes dispersionless, with compact
localized eigenstates and eigenenergies $E^-=\epsilon_n-\gamma$  which may partially overlap with 
$E^+$ (the overlapping depends on $\gamma$, see \cite{fb1} for details). 
A defining characteristics of mobility edges is the divergence of the localization length 
\begin{equation} 
\xi (E) \sim 1/(E - E_c)^{\nu}, \label{divergence}
\end{equation}
as a critical energy $E_c$ is approached \cite{mobility_edge}. Such a divergence is absent here, because \emph{all} the localized states are compact, $\xi( E) = 0$. Therefore this model does not host a mobility edge.
The case $t_d \ne t_l$ (Fig.3 in \cite{sil2008}) keeps the two chains noninteracting. None of the chains allows for a 
mobility edge, and therefore there is no mobility edge in the combined spectrum either.

The engineering of mobility edges in quasi-1D networks is indeed an interesting problem. 
This could be achieved by removing the symmetry and choosing $\epsilon_{n,1} \neq \epsilon_{n,2}$. This induces
a hybridization of the $f^+$ and $f^-$ states. At any given energy either only localized,  or only delocalized states 
can be expected, with several true mobility edges separating these classes of states.
\\
Acknowledgements
\\
We thank Daniel Leykam for useful discussions.

\end{document}